# First-principles predictions of out-of-plane group IV and V dimers as high-symmetry high-spin defects in hexagonal boron nitride

*Jooyong Bhang[1], He Ma[2,3], Donggyu Yim[1], Giulia Galli[2,3,4], and Hosung Seo[1]\**

[1]Department of Energy Systems Research and Department of Physics, Ajou University, Suwon, Gyeonggi 16499, Korea

[2]Pritzker School of Molecular Engineering, University of Chicago, Chicago, IL 60637, USA

[3]Department of Chemistry, University of Chicago, Chicago, IL 60637, USA

[4]Materials Science Division and Center for Molecular Engineering, Argonne National Laboratory, Lemont, IL 60439, USA

(\*Correspondence to hseo2017@ajou.ac.kr)

**ABSTRACT**

Hexagonal boron nitride (h-BN) has been recently found to host a variety of quantum point defects, which are promising candidates as single photon sources for solid-state quantum nanophotonics applications. Most recently, optically addressable spin qubits in h-BN have been the focus of intensive research due to their unique potential in quantum computing, communication and sensing. However, the number of high-symmetry high-spin defects that are desirable for developing spin qubits in h-BN is highly limited. Here, we combine density functional theory (DFT) and quantum embedding theories to show that out-of-plane $X_NY_i$ dimer defects (X, Y=C, N, P, Si) form a new class of stable $C_{3v}$ spin-triplet defects in h-BN. We find that the dimer defects have a robust $^3A_2$ ground state and $^3E$ excited state, both of which are isolated from the h-BN bulk states. We show that $^1E$ and $^1A$ shelving states exist and they are positioned between the $^3E$ and $^3A_2$ states for all the dimer defects considered in this study. To support future experimental identification of the $X_NY_i$ dimer defects, we provide extensive characterization of the defects in terms of their spin and optical properties. We predict that the zero-phonon line of the spin-triplet $X_NY_i$ defects lies in the visible range (800 nm – 500 nm) range. We compute the zero-field splitting of the dimers' spin to range from 1.79 GHz ($Si_NP_i^0$) to 29.5 GHz ($C_NN_i^0$). Our results broaden the scope of high-spin defect candidates that would be useful for the development of spin-based solid-state quantum technologies in two-dimensional hexagonal boron nitride.

**INTRODUCTION**

Optically addressable spin defects in wide-gap semiconductors[1] have recently shown immense promise for use in solid-state quantum information applications owing to their robust spin properties[2], spin-to-photon interfaces[3], and room-temperature functionality[4]. Furthermore, several vacancy defects in SiC[5,6] have featured prominently in the search for new and improved solid-state qubits compatible with scalable semiconductor devices[7] and telecommunication technologies[8]. Optically active quantum defects are also gaining prominence in integrated quantum photonics using two-dimensional materials[9]. Hexagonal boron nitride, a wide-band gap insulator among 2-dimensional (2D) van der Waals materials, has gained a great amount of attention due to the discovery of bright single-photon emitters operating at room temperature[10-14], and considerable focus has been devoted to the search of optically addressable spin qubits in h-BN.

Several breakthroughs have been recently achieved in the search of spin qubits in h-BN[15-17]. Notably, Gottscholl *et al.* demonstrated optical initialization and read-out of an ensemble of spin defects in h-BN, whose origin was suggested to be negatively charged boron vacancies ($V_B^-$)[15,18]. Motivated by the experimental observation[15], Ivady *et al.* and Reimers *et al.* constructed theoretical models of $V_B^-$ and showed that they can capture the essential magneto-optical features of the observed data[19,20]. Most recently, the deterministic generation of $V_B^-$ have been demonstrated[21] as well as coherent spin controls[22] making the $V_B^-$ system a leading spin qubit candidate in h-BN. In addition to $V_B^-$, Chejanovsky *et al.*, reported optically detected magnetic resonance (ODMR) experiments on single spins in h-BN and suggested carbon impurities as a possible origin[17]. More recently, Mendelson *et al.* observed room-temperature ODMR for ensembles of carbon-related defects in h-BN[23], confirming the potential of carbon-based defects as spin qubit candidates in h-BN[24-27]. Overall, great promises have emerged in the development of spin qubits in h-BN and their use in quantum information applications. Nonetheless, further research is required to realize spin qubits in h-BN with well-defined spin-to-photon interface and coherent single spin manipulation, which are akin to defect qubits in diamond and SiC.

Along with the experimental achievements, advances have been also made in theoretical defects-by-design approaches[28,29,30] to efficiently guide the exploration of spin qubits in h-BN[18,31,32]. A wide range of spin defects was theoretically considered, and several spin qubit candidates were suggested, including $V_B^-$[18], $V_BC_N$[32], and $Ti_{VV}$[31]. Notably, Abdi *et al.* predicted that optical spin initialization is possible in $V_B^-$ in h-BN owing to its non-trivial S=1 ground state and excited-state dynamics undergoing an intersystem crossing in the presence of $D_{3h}$ symmetry[18]. It is worth noting that the optical initialization mechanism of $V_B^-$ is essentially the same as those of other spin-triplet defect qubits in diamond and SiC[5,6].

Interestingly, however, besides $V_B^-$ or mono-impurities, the realization of defects with high-symmetry (e.g. $C_{3v}$) and high-spin (S > 1/2) is not trivial to achieve in h-BN[15,18,30], which may hinder the further discovery of optically active spin qubit candidates with superior quantum properties. In h-BN, the most studied defect structures for quantum applications are impurity-vacancy complexes, such as $V_NN_B$ and $V_NC_B$[10-12,18]. And the highest-possible symmetry of such complexes is $C_{2v}$ due to the 2-dimensional geometry of its host lattice[30]. It is worth noting that high-symmetry and high-spin are desirable for solid-state spin qubits owing to the many unique advantages that they can yield. Zero-field splitting (ZFS) parameters of high-spin defects are key quantities to develop nano-scale sensors to measure temperature[33], pressure[34], strain[35], and electric field[36]. ZFS parameters also enable coherent spin manipulation at zero-magnetic field and control using strain[37]. Orbital degeneracies with definite high-symmetry also play a crucial role in realizing spin-photon entanglement[38]. Therefore, it is highly desirable to explore the possibility of different types of high-symmetry defects in h-BN, which may lead to optically active defects with non-trivial spins.

In this study, we performed a series of first-principles calculations to show that out-of-plane $X_NY_i$ dimers (X, Y = C, N, P, Si) in h-BN are a class of new defects that are characterized by $C_{3v}$ symmetry ($D_{3h}$ for the $Si_NSi_i$ dimer) and spin-triplet ground state. Our density functional theory (DFT) results showed that the defects are energetically stable and exhibit a $^3A_2$ ground state localized in the band gap of hexagonal boron nitride. We predicted the optical zero-phonon line between the $^3A_2$ ground state and the $^3E$ excited state of the dimers to be ranging from 1.55 eV to 2.50 eV in the visible range. We found that the smallest and the largest Huang-Rhys factor of the $X_NY_i$ dimers are 4.47 for $S_NS_i^-$ and 23.02 for $C_NSi_i^-$, respectively. By using a quantum embedding theory[39] and exact diagonalization (QET-ED), we found that the dimer defects possess $^1E$ and $^1A$ singlet shelving states, which may play an important role in mediating intersystem crossing upon optical excitation. Overall, among the dimer models, we found that the $S_NS_i^-$ dimer may exhibits a preferable optical property as a high-spin defect candidate in h-BN. To guide future experimental identification of the dimers, we also predicted their spin Hamiltonian parameters including the hyperfine tensor and zero-field splitting parameter. Our results pave the way to developing new practical spin defect systems in h-BN, which would be beneficial for creating spin qubits for quantum information science and technology.

**THEORETICAL METHODS**

**Density functional theory calculations.** We performed DFT calculations as implemented in the Quantum Espresso (QE) code[40], using plane-wave basis sets with an energy cutoff of 85 Ry. To describe the interaction between core and valence electrons of B, N, C, Si, and P, we employed optimized norm-conserving Vanderbilt (ONCV) pseudopotentials[41] (SG15[42]) and projector-augmented wave (PAW) pseudopotentials[43] (QE PAW v0.3.1 set). We used the semi-local Perdew-Burke-Ernzerhof (PBE) functional and the hybrid Heyd-Scuseria-Ernzerhof functional (HSE)[44]. The Hartree-Fock mixing parameter ($\alpha$) in the HSE functional was set at a value of 0.40. The use of the HSE-40% functional to describe h-BN was proven to be accurate in several previous studies[27,45]. To describe the van der Waals interactions, present in multi-layer h-BN, we applied the non-local van der Waals (vdW) functional developed by Hamada (vdw-df2-b86r)[46] when carrying out PBE calculations, and the semi-empirical Grimme's DFT-D3 correction scheme[47] for the hybrid functional calculations. The ground-state structure of all the defects considered in this study was optimized by using the HSE-40% functional until the forces acting on the atoms are less than 10 meV/Å.

We calculated the defect formation energy (DFE) of C-based defects in multilayer h-BN at the hybrid functional level of theory by using the charge correction scheme developed by Freysoldt, Neugebauer, and Van de Walle[48]. The DFE of the $X_N Y_i$ dimer defects (X, Y = C, N, Si, P) other than the $C_N C_i$ defect was computed at the PBE level of theory. To identify the possible range of the chemical potentials in the DFE, we considered not only the energy of elementary bulk materials, but that of other potential competing phases such as $Si_3N_4$, BP, and $C_3N_4$[49]. We employed 240-atom supercells and we sampled the Brillouin zone (BZ) with the $\Gamma$-point only. A more detailed description of the calculation of the DFE can be found in our previous studies[50] and several review papers[28].

To calculate the zero-phonon line (ZPL) and the Huang-Rhys (HR) factor of the $X_N Y_i$ dimer defects, we carry out constrained DFT calculations[51] at the HSE level of theory. We computed the HR factor within the one-dimensional effective phonon method[52]. We note that the VASP code[53] was used for these excited-state calculations due to an improved performance in converging self-consistent hybrid DFT calculations in the excited states. For the VASP calculations, we used the same HSE-40% functional combined with the Grimme's DFT-D3 vdW correction scheme[47]. We employed the PAW pseudopotentials[43] along with a plane-wave cutoff energy of 700 eV. We modeled the XY dimer defects using a 240-atom supercell with the Γ-point BZ sampling.

We carried out calculations of hyperfine tensors and zero-field splitting tensors of the dimer defects in h-BN by using the GIPAW module of the QE code[54] and the PyZFS code[55], respectively, at the PBE level of theory. In the case of hyperfine tensor calculations, we used the PAW pseudopotentials and we included core polarization effects in the evaluation of spin densities near nuclei[56]. For the D-tensor calculations, we adopted the method proposed by Rayson and Briddon[57] and evaluated the ZFS tensor using the Kohn-Sham orbitals obtained with ONCV pseudopotentials. We used 256-atom supercells and the Γ-point only. We note that the use of the PBE functional for spin Hamiltonian parameter calculations were verified for a wide range of spin defects in solids[29,58], thus providing a reliable ground for our predictions. We remark, however, that a comprehensive comparison with calculations based on hybrid DFT would be necessary and informative[59] and it will be the topic for future work.

**Quantum embedding theory and exact diagonalization (QET-ED).** Intra-defect transitions of the $X_N Y_i$ dimers lead to a rich set of electronically excited states, many of which are expected to be strongly correlated in nature, based on molecular orbital theory and their localization properties; therefore, they are not directly accessible by Kohn-Sham DFT calculations[60]. We applied the QET developed in Ref.[39] to construct effective Hamiltonians acting on an active space including selected defect levels in the band gap. The effective Hamiltonian is solved by exact diagonalization (ED), yielding the vertical excitation energies of the system.

We constructed the effective Hamiltonian based on spin-restricted DFT calculations of defects using the HSE functional and a 129-atom supercell along with the Γ-point only. The relatively small supercell was employed to reduce the computational cost of subsequent embedding theory calculations, while insuring a correct description of the dimer defects' electronic structure. We note that the defect wavefunctions are highly localized in space, and therefore the Kohn-Sham eigenvalues and the wavefunctions used by the embedding theory calculations are well reproduced with the 129-atom supercell. We performed convergence tests by considering various supercell sizes and the results are summarized in Supplementary Table S1. We found that the 129-atom supercell calculation yields excited-state energies converged within ~0.15 eV. The effective electron-electron interaction in the Hamiltonian is evaluated from a partially screened Coulomb interaction where the density response within the active space is projected out. The partially screened Coulomb interaction is computed using the random phase approximation (RPA) on the space spanned by the first 512 eigenvectors of the irreducible density response function of the dimers using the projective dielectric eigen-decomposition (PDEP) technique[61,62]. It is worth mentioning that beyond-RPA calculations combined with the hybrid functional would be informative to better understand the excitation energies and will be the topic of future studies. The effective Hamiltonian is constructed using an active space that includes the $a$ and $e$ levels of the defects that fall in the band gap of h-BN, in addition to the valence band states within 3 eV below the valence band maximum. The QET calculations are performed with the WEST code[62,63,64]. The exact diagonalization of the effective Hamiltonian is carried out with the PySCF code[65].

**RESULTS**

**Ground-state properties of the spin-triplet $X_NY_i$ dimer defects in h-BN.** We begin by discussing the ground-state properties of out-of-plane C-C dimer defects as shown in Fig. 1(a) as a representative system of the X-Y dimer defects in h-BN (X, Y = C, Si, P, N). In general, for out-of-plane C-C dimer defects, there are two possible positions in the h-BN lattice: $C_NC_i$ and $C_BC_i$, in which C substituting for N ($C_N$) and B ($C_B$), respectively, pairs with an interstitial C impurity ($C_i$). However, we find that the $C_BC_i$ defect does not exhibit a ground-state spin higher than ½ in any of its possible charge states, but the $C_NC_i$ defect obtains a spin-triplet ground state along with $C_{3v}$ symmetry in its negative charge state (i.e. $C_NC_i^-$). As our study aims at identifying defects in h-BN, whose spin and symmetry are higher than ½ and $C_{2V}$, respectively, we focus on the $C_NC_i^-$ defect. In what follows, we describe the ground-state property of the $C_NC_i^-$ defect in detail.

Figure 1(a) also shows the optimized ground-state structure of the $C_NC_i^-$ defect, in which the dimer defect is bonded to four nearest neighboring B atoms forming a $C_{3v}$ trigonal pyramid geometry. The $C_N$-$C_i$ bond length is calculated to be 1.47 Å, which is larger than the bond length of the $C_2$ molecule in a vacuum (1.24 Å), but similar to that in graphite (1.42 Å). The deviation of 0.23 Å from the $C_2$ molecule bond length may be due to the chemical bonding interaction between the $C_NC_i^-$ dimer and the four nearest neighboring B atoms. Besides, the apical B atom which is the nearest neighbor of the $C_i$ atom in the dimer is significantly displaced by 0.42 Å off the layer toward the $C_NC_i^-$ defect due to the electrostatic attraction between the positive B ion and the negatively charged $C_NC_i^-$ defect. To cross-check the stability of the $C_{3v}$ S=1 state, we also computed various distorted structures in a low-spin S=0 state (see Supplementary Figure S1). We find, however, that the $C_{3v}$ S=1 state is lower in energy by 250 meV than the lowest-energy S=0 state in a $C_{1h}$ structure at the HSE level of theory, corroborating the robustness of the high-symmetry high-spin ground-state of the $C_NC_i^-$ defect.

To shed light on the microscopic origin of the spin-triplet ground state of the $C_NC_i^-$ defect, we analyze the electronic structure of the $C_NC_i^-$ defect. Figure 1(b) shows the defect level diagram of the $C_NC_i^-$ defect, which corresponds to a $^3A_2$ spin-triplet state. Degenerate $e$ states are found at 4.1 eV in energy below the conduction band minimum (CBM), and they are occupied by two unpaired electrons leading to the spin-triplet state. Also, a fully occupied $a$ state is present near the valence band maximum (VBM): 0.35 eV and 0.58 eV above the VBM in the spin up and down channels, respectively.

We find that the defect level diagram of the $C_NC_i^-$ defect can be qualitatively understood by using molecular orbital theory. We consider the molecular orbitals of a $C_2$ dimer in a $C_{3v}$ environment as shown in Fig. 1(c). Among the six $p$ orbitals of the two C atoms, the $p_x$ and $p_y$ orbitals form fully occupied $2p\pi$ bonding orbitals and empty $2p\pi^*$ anti-bonding orbitals. Furthermore, the two $p_z$ orbitals create empty $2p\sigma$ and $2p\sigma^*$ orbitals, which are below and above the $2p\pi^*$ state, respectively. When the $C_2$ molecule is brought into h-BN as $C_NC_i$ as the one shown in Fig. 1(a), the $2p\pi^*$ and $2p\sigma$ orbitals transform as $e$ and $a_1$ irreducible representations, respectively. In the ionic limit, each of the three nearest neighboring B atoms may transfer one electron to $C_NC_i$. Therefore, in its negative charge state, $C_NC_i$ in h-BN would acquire four extra electrons compared to the bare $C_2$ molecule in a vacuum, and the $2p\pi^*$ $e$ manifold would be occupied by two unpaired electrons. We observe that this theory is fully consistent with the defect level diagram shown in Fig. 1(b). In addition, the $a$ and $e$ defect orbitals shown in Fig. 1(d) exhibit $2p\sigma$ and $2p\pi^*$ orbital character, respectively, as predicted by molecular orbital theory. Our results show that the out-of-plane $C_NC_i^-$ dimer attains the $^3A_2$ spin-triplet ground-state due to the bonding and anti-bonding interaction of the C $2p$ orbitals and the electron transfer from h-BN.

Inspired by the result of $C_NC_i^-$, we generalize the idea of $C_NC_i^-$ by including other group-IV and group-V elements as dimer's constituent atoms, including Si, P, and N. Surprisingly, we find that all the $X_NY_i$ dimer defects (X=C, N, Si, and P) considered in this study give rise to high-symmetry ($C_{3v}$ or $D_{3h}$) spin-triplet ground-state when appropriately charged. Figure 2 shows five $X_NY_i$ dimer defects in a spin-triplet ground-state out of ten possible combinations. We find that $C_NN_i$, $C_NP_i$, and $Si_NP_i$ have a spin-triplet ground state in their neutral charge state, while a negative charge state ($q$ = -1) is required for $C_NSi_i$ and $Si_NSi_i$. We only show those five cases because the atomic structure of a few other possible $X_NY_i$ dimer defects converged to the same dimer structures shown Fig. 2 upon DFT structural relaxation. For instance, we prepared ideal initial structures of $P_NC_i^0$ and $Si_NC_i^-$, but these relaxed to the structure of $C_NP_i^0$ and $C_NSi_i^-$, respectively, shown in Fig. 2(b) and 2(c). $N_NN_i$ is also not considered as this is the same as an interstitial N impurity. In terms of the dimer geometry, it is interesting to observe that the Si atom in the $Si_NP_i^0$ dimer and $Si_NSi_i^-$ dimer is off the h-BN plane as shown in Fig. 2(d), and 2(e), respectively, which may be due to the large atomic size of Si and P. As a result, the $Si_NSi_i^-$ defect has the highest-symmetry of $D_{3h}$ and other dimer defects possess the $C_{3v}$ symmetry.

Fig. 2 also shows the defect level diagram of the spin-triplet $X_NY_i$ defects. For all of the dimer defects, the highest-occupied defect levels are doubly degenerate $e$ orbitals being occupied by two unpaired electrons, which corresponds to a $^3A_2$ ground state. By considering the in-gap defect orbitals shown in Supplementary Figure S2, we find that these highest-occupied $e$ orbitals are mainly derived from the $p\pi^*$ molecular orbitals of the dimer's constituent atoms as found in the $C_NC_i^-$ defect's electronic structure. However, there are also differences depending on the dimer type and charge state. While the main feature of the $C_NSi_i^-$'s defect level diagram is the same as that of the $C_NC_i^-$, only the $e$ orbitals are visible in the band gap for $C_NN_i^0$ and $C_NP_i^0$, but more complex in-gap level structures are found for $Si_NP_i^0$ and $Si_NSi_i^-$. For $C_NN_i^0$ and $C_NP_i^0$, we find that the $a_1$ orbital associated with the $p\sigma$ interaction is formed below the VBM. For $Si_NP_i^0$ and $Si_NSi_i^-$, we note that interlayer coupling between the dimer and the adjacent h-BN layer plays an important role in producing such complex level structure as indicated by the molecular orbital shape shown in Supplementary Figure S2.

To test the stability of the spin-triplet ground state of the $X_NY_i$ dimers, we used the same DFT total energy method that we applied to the $C_NC_i^-$ defect. In Supplementary Table S2, we compare the energy of the high-symmetry high-spin state of the $X_NY_i$ dimers to the lowest-energy of various distorted S=0 states of the dimers at the HSE level of theory. We find that for all the dimer defects considered in this study, the spin-triplet state is always lower in energy by more than 100 meV than the low-symmetry spin-singlet state. Our results show that the $X_NY_i$'s high-spin ground-state is robust.

**Defect's charge-state stability.** We remarked that the spin-triplet ground state of the $X_NY_i$ dimers is achieved when the defects are appropriately charged, as summarized in Fig. 2. Thus, it is crucial to examine that the required charge state for each dimer is indeed stable. Again, taking the $C_NC_i$ dimer as a representative system of the $X_NY_i$ dimers, we compute the defect formation energy (DFE) of the $C_NC_i$ defect at the HSE level of theory. As shown in Fig. 3, we find that the negative charge state is stable for the $C_NC_i$ defect when the Fermi level ranges from 3.43 eV ((0/-1) charge transition level (CTL)) to 4.83 eV ((-1/-2) CTL)). Other possible charge states are found to be +1, 0, and -2. By taking the difference between the conduction band minimum (CBM) and the (0/-1) CTL, we obtain the ionization energy of the negatively charged $C_NC_i$ defect, which is 3.02 eV. We note that the ionization energy is larger than the predicted ZPL of 2.48 eV, indicating that the negative charge state of the $C_NC_i$ defect is stable under one-photon optical excitation and emission cycle near the ZPL wavelength[28].

To investigate the relative stability of the $C_NC_i$ defect, we compare the DFE of the $C_NC_i$ defect to those of other C-derived impurities that can be easily incorporated in h-BN[23,25], including interstitial C ($C_i$), C substituting for N ($C_N$), and C substituting for B ($C_B$). Fig. 3(a) and 3(b) show the DFE of the defects in the N-poor and N-rich limits, respectively. For the $C_N$, $C_B$, and $C_i$ defects, we find that our results are in good agreement with previous results reported by Weston *et al.*[66].

Our results in Fig. 3(a) show that in the presence of both $C_N$ and $C_i$ in h-BN, it is energetically favorable to form the $C_NC_i$ dimer structure rather than separate entities. In both Fig. 3(a) and 3(b), the DFE of $C_i$ is independent of the N and B chemical potentials as the formation of $C_i$ does not involve N or B. Thus, the DFE of the $C_NC_i$ or $C_BC_i$ dimer largely depends on its constituent substitutional impurity. In the N-poor limit shown in Fig. 3(a), the DFE of $C_N$ is lower than that of the $C_B$ by more than 2 eV, leading to a low DFE of $C_NC_i$, which is even smaller than those of $C_i$ and $C_BC_i$. In the N-rich limit shown in Fig. 3(b), however, the tendency is opposite: the DFE of $C_BC_i$ is smaller than those of $C_NC_i$ and $C_i$ as the DFE of $C_B$ is lowered compared to that of $C_N$. Thus, our results suggest that the realization of the $C_NC_i$ defects in their negative charge state in h-BN could be achieved in the N-poor limit in the presence of both $C_i$ and $C_N$ in the sample. We also find that the DFE of the $C_NC_i^-$ defect is comparable to those of other well-known intrinsic and extrinsic defects in h-BN[66], indicating that the experimental realization of the defect may be possible.

Motivated by the DFE result for the $C_NC_i$ dimer defect, we compute the DFE of the other $X_NY_i$ dimer defects at the PBE level of theory and the results are summarized in Supplementary Figure S3. By comparing the DFE of $C_NC_i$ in Figure S3 to that in Fig. 3, we check that the PBE DFE qualitatively agrees with the DFE computed at the more accurate and computationally expensive HSE level of theory, verifying its usage in understanding the charge state stability of the $X_NY_i$ dimer defects other than $C_NC_i$. As highlighted in Fig. S3, we observe that the required charge state enabling the high-symmetry high-spin ground state for the $X_NY_i$ dimers is stable: neutral charge state for $C_NN_i$, $Si_NP_i$, and $C_NP_i$, and negative charge state for $C_NSi_i$, $C_NC_i$, and $Si_NSi_i$. In addition, we observe that the DFE of the $X_NY_i$ dimer defects in Figure S3 is also comparable to those of known intrinsic defects in h-BN[66] in the nitrogen-poor limit, suggesting that the experimental creation of the defects may be achievable.

To better understand the potential process of $C_NC_i^-$ dimer formation, we compute the potential energy surface of the defect as a function of the $C_i$'s (x,y) coordinate in the presence of nearby $C_N$ and the results are presented in Fig. 4. We find that $C_N$ acts as a global trapping site for $C_i$. As shown in Fig. 4(b), local trapping sites for $C_i$ away from the $C_N$ site are either B or N sites. However, the energy barrier between the local trapping sites is estimated to be as low as 0.2 eV, indicating that $C_i$ is not strongly bound by these local trapping sites. We find, however, that a strong chemical bonding can occur between $C_i$ and $C_N$. In our calculations, the total energy is significantly lowered when $C_i$ is bound to the $C_N$ site. The strong binding between the $C_N$ and $C_i$ is due to the bonding interaction between the C 2p states as described in Fig. 2. Besides, we observe in Fig. 4(b) that the trapping site for the $C_i$ defect includes not only the $C_N$ site but extends to the second n.n. N sites of the $C_N$ site. The large trapping area around the $C_N$ defect would be beneficial for the formation of the $C_NC_i$ dimer defect in h-BN.

For the spin-triple $X_NY_i$ defects other than $C_NC_i^-$, we compute the binding energy gain by computing the total energy difference between the ground-state energy of the spin-triplet $X_NY_i$ dimer, and the energy of a defect configuration, in which the X atom substitutes for N ($X_N$), but the Y atom is placed at its lowest-energy interstitial site away from the X atom. The results are summarized in Supplementary Table S3. We find that for all the spin-triplet $X_NY_i$ dimer defects considered the energy gains in forming dimers are larger than 1 eV.

**Excited-state properties of the spin-triplet $X_NY_i$ dimers in h-BN.** The optical initialization and readout of the diamond NV center and the SiC defect qubits relies on the spin conserving excitation to an $^3E$ excited state and subsequent spin-selective decay to the $^1E$ and $^1A_1$ spin-singlet shelving states[1]. Therefore, it is important to understand the many-electron excited states of the spin-triplet $X_NY_i$ dimers to examine their potential as optically addressable spin qubit candidates in h-BN.

We perform QET+ED calculations using the HSE ground-state structure and wavefunctions, and the results are summarized in Table 1. Remarkably, our results show that the spin-triplet $X_NY_i$ defects exhibit an excited-state structure that is qualitatively the same as that of the diamond NV center[67]. We find that a spin-conserving vertical excitation can occur between the $^3A_2$ ground state ($a^2e^2$ configuration) and the $^3E$ excited state ($a^1e^3$ configuration) and $^1A$ and $^1E$ shelving states are also present between the $^3A_2$ and the $^3E$ states for all the spin-triplet $X_NY_i$ dimer defects considered. In particular, we observe that the $^1A$ state of the $Si_NSi_i^-$ defect is close in energy to the $^3E$ state within 0.5 eV, which may play a significant role in intersystem crossing.

To investigate the possibility of intersystem crossing upon optical excitation, we also compute the many-electron multiplet structure of the spin-triplet $X_NY_i$ dimers at their excited-state geometry (see Supplementary Figure S4), which is summarized in Supplementary Figure S5. We find that at the excited-state geometry the ordering of the $^3E$ triplet states and the $^1A$ and $^1E$ singlet states are reversed. This result suggests that the $^1A$ and $^1E$ states may mediate an intersystem crossing, which could lead to optical initialization of the $X_NY_i$ spins. We remark, however, that to fully confirm the possibility of optical spin initialization in the spin-triplet $X_NY_i$ dimers, further analysis should be carried out by investigating their optical property and excited-state dynamics in detail, including oscillator strengths and intersystem crossing rates, which will be the subject of a future study.

Table 2 summarizes the ZPL energies and the HR factors of the spin-triplet $X_NY_i$ dimer defects at the HSE level of theory computed by using the VASP code. We used a constrained DFT method to compute the ZPL energy by promoting an electron from the *a* orbital to the *e* orbital and performing geometry relaxation. The ZPL energies and the HR factors computed at the PBE level of theory are reported in Supplementary Table S4. We found that the ZPL energies range from 1.55 eV ($Si_NSi_i^-$) to 2.50 eV ($C_NP_i^0$), and thus belong to the visible spectrum.

Interestingly, we observe that the HR factors of the spin-triplet $X_NY_i$ dimers bearing $C_N$ are ~22, while those of $Si_NP_i^0$ and $Si_NSi_i^-$ are 14.81 and 4.47, respectively. We find that the $C_N$-based dimer defects undergo a significant structural relaxation in the excited state (see Supplementary Figure S6 (a to d)), leading to the large HR factors. This is mainly because the $Y_i$ atom (Y = C, N, P, Si) in the $C_N$-bearing $X_NY_i$ dimers is placed between h-BN layer, and thus it is surrounded by a large space between the h-BN planes. A significant structural relaxation of the $Y_i$ atom also induces a noticeable out-of-plane distortion in the adjacent h-BN layer. In contrast to the $C_NY_i$ (Y = C, N, P, Si) dimers, both the $Si_N$ and $Z_i$ atoms in the $Si_NZ_i$ (Z = P and Si) dimer are strongly bound to the h-BN layer embedding the dimer (see Supplementary Figure S6 (e, f)). In the excited state, the $Si_NZ_i$ (Z=P and Si) defect undergoes minor structural relaxations, leading to a smaller HR factor than those of the $C_NY_i$ (Y = C, N, P, Si) dimers.

**Spin Hamiltonian parameters.** Electron paramagnetic resonance (EPR) is one of the essential techniques to characterize defect spins. Spin Hamiltonians describing the interactions of an electronic spin with nuclear spins play a key role in the analysis of EPR signals. For a $C_{3v}$ spin-triplet defect, spin Hamiltonians include zero-field splitting (ZFS) and hyperfine terms[29,67]. The former is described by two parameters, the axial (D) and rhombic (E) parameters, and in the case of a spin-triplet defect, the axial D parameter describes a splitting between the $m_s = \pm 1$ and $m_s = 0$ spin sub-levels even in the absence of external magnetic field.

Table 3 summarizes the computed D parameters of the spin-triplet $X_N Y_i$ dimers and the D value ranges from 1.79 GHz for $Si_N P_i^0$ to 29.5 GHz for $C_N N_i^0$. We note that the D parameter decreases as the dimer's constituent atoms contain more $3p$ electrons. This is because the D parameter is originated from the dipolar spin-spin interaction[55,57]. In the $C_N C_i^-$ dimer, its spin density is mainly derived from the carbon $2p$ orbitals, which are highly localized in space. In contrast, the spin density in the $Si_N P_i^0$ dimer is more delocalized in space compared to that of the $C_N C_i^-$ dimer as it is derived from the $3p$ orbitals of Si and P. We also note that we only consider the spin-spin interaction while the spin-orbit contribution to the ZFS parameters is not included, which may underestimate the D parameter, in particular of the Si- and P-bearing dimer defects.

The hyperfine tensors describe the interaction between an electron spin (S) and nuclear spins (I). In h-BN, there are two naturally occurring B isotopes with non-zero nuclear spins, $^{10}$B (I=3, natural abundance of 80.1%) and $^{11}$B (I=3/2, 19.9 %). N also has two different isotopes possessing intrinsic nuclear spins, $^{14}$N (I=1, 99.6%) and $^{15}$N (I=1/2, 0.37%). For C, $^{13}$C isotopes exhibit nuclear spin ½ at 1.1% natural abundance, while $^{12}$C is nuclear-spin free. In Table 4, we report the hyperfine parameters of the spin-triplet $X_NY_i$ dimer's electron spin and the nearest neighboring nuclear spins as these are the largest hyperfine parameters, which would dominate the hyperfine signal of the $X_NY_i$ defects. We find that for the $C_N$-bearing $X_NY_i$ dimers the interstitial $Y_i$ atom and the basal B atoms exhibit large contact hyperfine terms due to the highly localized spin density at the $Y_i$ site and the B sites. For $Si_NP_i^0$ and $Si_NSi_i^-$, however, we notice that the contact hyperfine parameters listed in Table 4 are reduced below 10 MHz mainly due to the delocalized spin density beyond the dimer and its nearest neighboring sites. We also computed the anisotropic dipolar interactions and their principal axes, and these results are reported in Supplementary Table S5. We find that the magnetic anisotropy axis of the spin-triplet $X_NY_i$ defects is parallel to the $X_N$-$Y_i$ dimer direction, consistent with the $C_{3v}$ symmetry of the defects.

**CONCLUSIONS**

In summary, we used first-principles calculations to show that the $X_NY_i$ dimer defects (X, Y = C, N, Si, and P) are a new class of $C_{3v}$ spin-triplet defects in h-BN. We showed that the spin-triplet $X_NY_i$ defects acquire robust $C_{3v}$ symmetry ($D_{3h}$ for the $Si_NSi_i^-$ dimer) by out-of-plane chemical bonding of interstitial $Y_i$ to substitutional $X_N$. Their doubly degenerate *e* orbitals localized in the band gap of h-BN are occupied by two unpaired electrons, leading to a $^3A_2$ ground state and a $^3E$ excited state. We found that the optical zero-phonon line for the $^3A_2 – ^3E$ spin-conserving excitation of the spin-triplet $X_NY_i$ dimers range from 1.55 eV to 2.50 eV, which are in the visible range. Among the $X_NY_i$ dimers, we found that the $Si_NSi_i^-$ dimer has the smallest Huang-Rhys factor of 4.47, which is comparable to that of the diamond NV center[1]. We also demonstrated that the defects possess $^1A$ and $^1E$ spin-singlet shelving states between the $^3E$ and $^3A_2$ states. For the $Si_NSi_i^-$ defect, the $^1A$ singlet state is found at 0.5 eV in energy below the $^3E$ state. We predicted the existence of level crossings between the shelving states and the $^3E$ excited state upon optical excitation in the presence of phonons, which may be beneficial to realize optical initialization and readout based on the intersystem crossing. Overall, among the $X_NY_i$ dimer defects, we found that the $Si_NSi_i^-$ defect may exhibit a preferable optical property as a spin defect candidate, which is similar to the diamond NV center[1].

However, we note that further investigations are necessary in order to fully support the potential intersystem-crossing-driven optical initialization of the spin-triplet $X_NY_i$ defects[19,20,68]. The optical spin initialization could occur in the presence of a spin-selective decay, which depends on competition between radiative and non-radiative decays upon optical excitation. In this complex process, the spin-orbit coupling, and Jahn-Teller effects may play an important role since the transition between the $^3E$ excited state and the singlet shelving state is mediated by the spin-orbit coupling between these states and phonons[69]. These are, however, beyond the scope of this study and will be the topic of separate future investigations.

To guide experimental investigations, we reported the zero-field splitting parameters and the hyperfine parameters of the spin-triplet $X_NY_i$ defects. We found that a large zero-field splitting parameter (D) ranging from 1.8 GHz for $Si_NSi_i^-$ to 29.5 GHz for $C_NN_i^0$. We remark that these frequency ranges could be covered by conventional Q-band or X-band electron paramagnetic resonance (EPR) methods. In addition, the large zero-field splitting parameter of the spin-triplet $X_NY_i$ dimers would be advantageous to decouple the defect spin from background spin-1/2 paramagnetic impurities. Further, the large zero-field splitting would help elongate the relaxation time of the spin by suppressing flip-flop transitions between the defect spin and bath spins.

It is worth mentioning that recent experimental studies reported C-related optically active defect centers[25] and Si impurities[70] in h-BN. Also, some of them showed single photon emissions and ODMR signals[23], indicating the possibility of C- or Si-related optically active spin defects. We remark, however, that the characteristics of the newly suggested dimer models obtained in this study may not correspond to the experimentally characterized emissions. We note that more detailed understanding would be certainly required to realize the newly suggested C- or Si-related spin-triplet dimers in h-BN in experiments.

The spin-triplet $X_NY_i$ models developed in this study has implications beyond a single defect type in h-BN. The essential design concepts could be applied not only to different interstitial defects in h-BN but to an exploration of spin defects in a wide range of two-dimensional materials. Interstitial impurities are ubiquitous in any 2-dimensional crystals. They often exhibit highly localized deep-level states in the band gap. However, oftentimes they are highly mobile, making them structurally unstable. Our defect design scheme paves the way to the realization of robust high-symmetry, high-spin quantum defects based on interstitial defects bound to substitutional defects, which would be favorable for the development of spin-based quantum applications in two-dimensional materials.

**FIGURE**

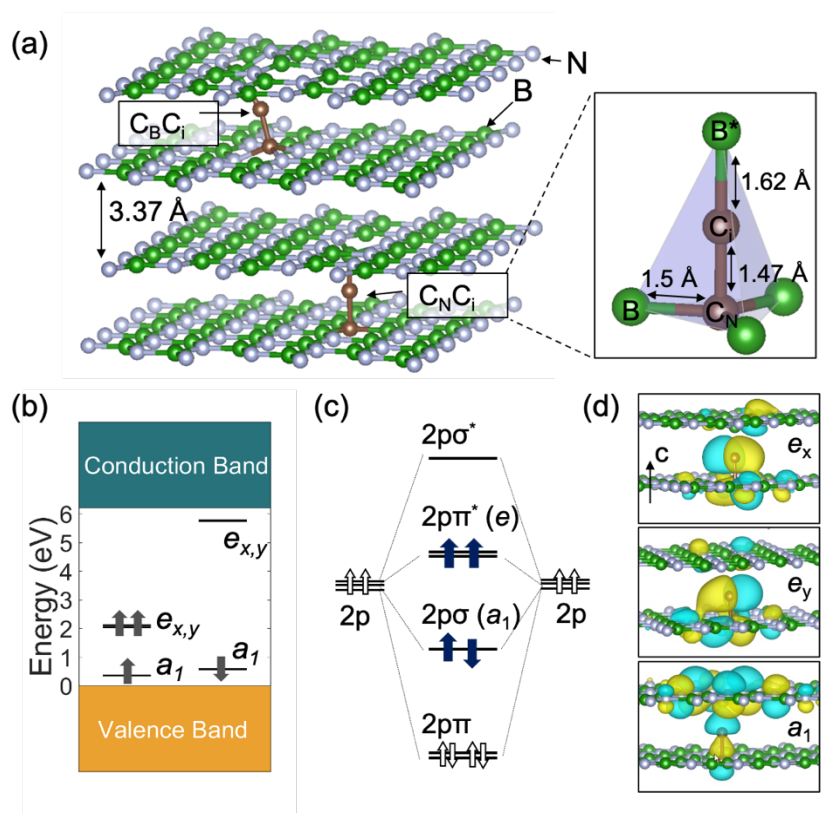

**FIG. 1.** (a) Possible atomic structures of out-of-plane negatively charged carbon dimers ($C_NC_i^-$ and $C_BC_i^-$) in hexagonal boron nitride. (gray sphere = nitrogen, green sphere = boron, brown sphere = carbon) An enlarged schematic of the $C_NC_i^-$ atomic structure with its four nearest neighboring B atoms is shown on the right. (b) Defect level diagram of the $C_NC_i$ defect in its negative charge state calculated at the hybrid functional level of theory. (c) Molecular orbital diagram of a $C_2$ dimer in a vacuum with four extra electrons, which are indicated by filled arrows. (d) Charge density (wavefunction squared) associated with the $a$ and $e$ defect orbitals shown in Fig. 1(b). The phase of the wavefunction is indicated by a different color (yellow for positive and blue for negative).

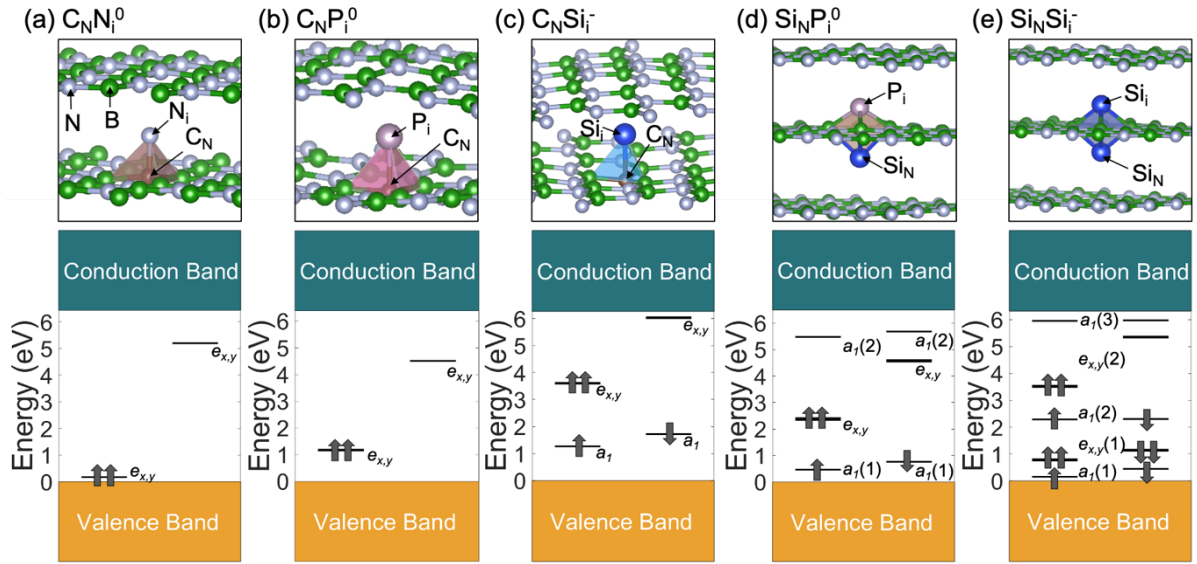

**FIG. 2. (a – e)** The atomic structure (top) and the defect level diagram (bottom) of five $X_N Y_i$ dimers (X, Y = C, P, Si, N) exhibiting the $C_{3v}$ symmetry ($D_{3h}$ for $Si_N Si_i^-$) and the spin-triplet ground state, computed at the HSE level of theory. The charge state of each defect is denoted in the superscript: 0 and – for the neutral and negatively charge states, respectively. In the defect level diagram, electrons occupying the defect orbitals are denoted as up-arrows and down-arrows in the spin-up and spin-down channels, respectively. For the $Si_N Si_i$ defect, the defect orbitals are labeled with $C_{3v}$ irreducible representations (*a* and *e*) although the defect symmetry is $D_{3h}$ for consistency with other dimer defects.

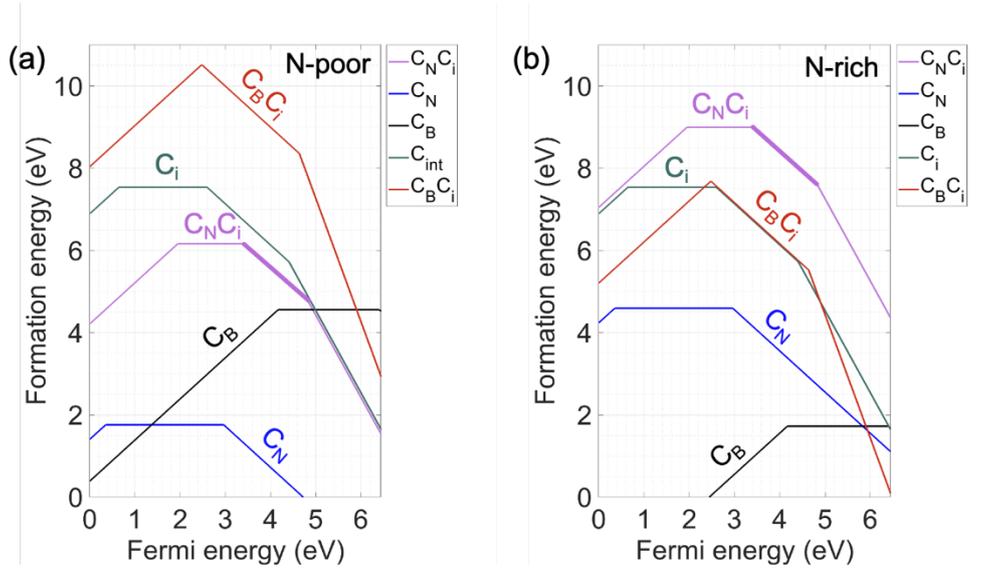

**FIG. 3.** (a, b) Defect Formation energy of carbon-related defects in h-BN in the nitrogen-poor (a) and nitrogen-rich (b) condition calculated at the hybrid density functional level of theory. The following defects are considered: C substituting for N ($C_N$, blue line), C substituting for B ($C_B$, black line), interstitial C placed between h-BN planes ($C_i$, green line), $C_NC_i$ defect pair (purple line), and $C_BC_i$ defect pair (red line). For the $C_NC_i$ defect, the spin-triplet ground state is found in the negatively charged ($q = -1$) state, which is denoted by a thick solid line in its defect formation energy plot.

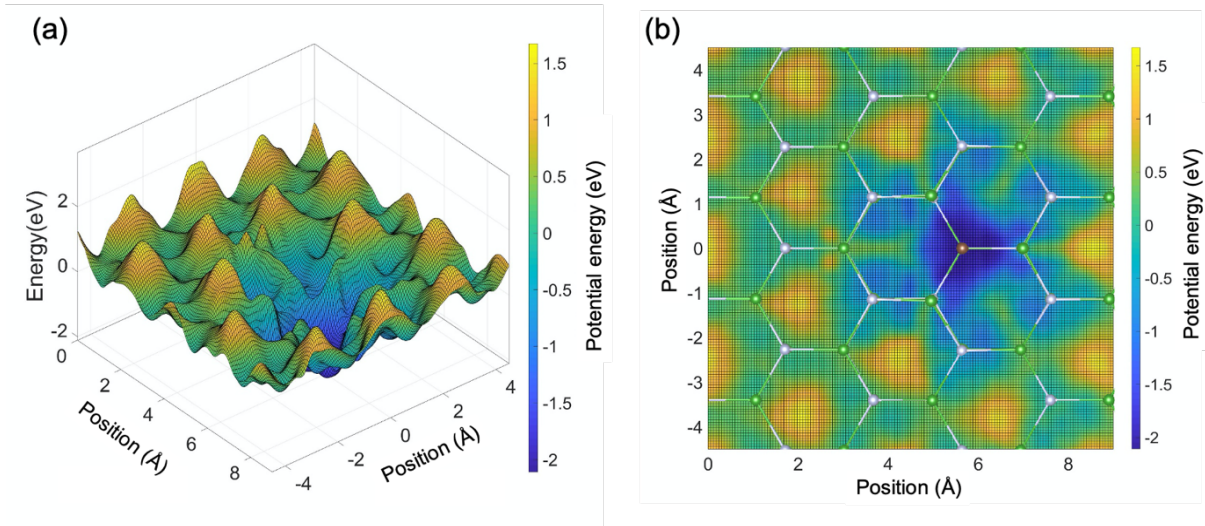

**FIG. 4. (a,b)** Potential energy surface (PES) of the negatively charged $C_NC_i$ defect as a function of the (x,y) position of the $C_i$ atom in the presence of C substituting for N ($C_N$) at a fixed position in a three-dimensional plot (a) and in a two-dimensional plot (b). To guide eyes, the structure of h-BN including $C_N$ is overlaid in (b): the brown sphere indicates $C_N$, green and grey spheres denote B and N atoms, respectively. In order to calculate the PES, the lowest energy structure of $C_i$ is calculated for a given (x,y) position with respect to the $C_N$ defect.

.

**TABLE 1.** Many-electron excited states of the $X_N Y_i$ dimers computed by using QET+ED. The $^3A_2$ ground state energy is set to be 0 eV.

| States | $C_N C_i^-$ (eV) | $C_N N_i^0$ (eV) | $C_N P_i^0$ (eV) | $C_N Si_i^-$ (eV) | $Si_N P_i^0$ (eV) | $Si_N Si_i^-$ (eV) |
|---|---|---|---|---|---|---|
| $^3A_2$ | 0.000 | 0.000 | 0.000 | 0.000 | 0.000 | 0.000 |
| $^1E$ | 0.505 | 0.708 | 0.520 | 0.407 | 0.320 | 0.263 |
| $^1A$ | 1.063 | 1.375 | 1.038 | 0.811 | 0.871 | 0.741 |
| $^3E$ | 2.824 | 2.176 | 2.185 | 2.218 | 1.643 | 1.233 |
| $^1E$ | 3.040 | 2.334 | 2.575 | 2.690 | 2.014 | 1.485 |

**TABLE 2.** Computed zero-phonon line (ZPL) and Huang-Rhys (HR) factor of the $X_N Y_i$ dimers in h-BN for the $^3A_2 - {}^3E$ transition at the HSE level of theory computed by using the VASP code. The one-dimensional effective phonon approximation was adopted to compute the ground-state phonon frequency ($\Omega$), the configurational coordinate change ($\Delta Q$) between the ground and the excited states, and the modal mass ($M$).

| Defect | ZPL (eV) | HR factor | $\Omega$ (THz) | $\Delta Q$ ($\sqrt{a.m.u}$Å) | $M$ (a.m.u) |
|---|---|---|---|---|---|
| $C_N C_i^-$ | 2.481 | 22.65 | 57.21 | 2.240 | 3.481 |
| $C_N N_i^0$ | 2.266 | 22.65 | 38.45 | 2.733 | 3.624 |
| $C_N P_i^0$ | 2.496 | 20.56 | 60.40 | 2.077 | 4.009 |
| $C_N Si_i^-$ | 2.290 | 23.02 | 40.30 | 2.691 | 3.919 |
| $Si_N P_i^0$ | 1.917 | 14.81 | 33.39 | 2.372 | 3.623 |
| $Si_N Si_i^-$ | 1.548 | 4.47 | 60.03 | 0.972 | 3.545 |

**TABLE 3.** Computed zero-field splitting parameters of the $X_NY_i$ dimers obtained at the PBE level of theory.

| Defect | D (GHz) |
|--------|---------|
| $C_NN_i^0$ | 29.5 |
| $C_NC_i^-$ | 13.6 |
| $C_NP_i^0$ | 6.05 |
| $C_NSi_i^-$ | 3.19 |
| $Si_NSi_i^-$ | 2.38 |
| $Si_NP_i^0$ | 1.79 |

**TABLE 4.** Computed isotropic hyperfine parameters (Fermi contact) of the $X_NY_i$ dimer defects calculated at the PBE level of theory. Hyperfine interactions with the nuclear spins of the nearest neighboring B atoms are also considered. The boron atoms which are the nearest neighbors of $X_N$ are denoted as B in the table, while the apical B* atom which is the nearest neighbor of $Y_i$ is denoted as B*. Information about the principal axes can be found in Supplementary Information.

| Defect | B (MHz) | B* (MHz) | $X_N$ (MHz) | $Y_i$ (MHz) |
|---|---|---|---|---|
| $C_NC_i^-$ | 13.0 | -16 | 5.5 | 46.7 |
| $C_NN_i^0$ | 69.9 | -14.7 | -0.8 | 54.9 |
| $C_NP_i^0$ | 21.2 | 1.3 | 3.0 | 25.4 |
| $C_NSi_i^-$ | 12.1 | -0.4 | -0.8 | -19.1 |
| $Si_NP_i^0$ | 7.7 | -0.4 | -6.1 | -0.9 |
| $Si_NSi_i^-$ | 6.3 | -1.3 | -5.5 | -5.5 |

## ASSOCIATED CONTENT

**Supporting Information.** Supplementary Table S1. Results of numerical convergence tests of the QET-ED calculations. Supplementary Figure S1. Atomic structure and electronic structure of the low-spin (S=0) $C_NC_i^-$ defect. Supplementary Figure S2. Charge density (wavefunction squared) associated with the *a* and *e* defect orbitals of the $X_NY_i$ dimers. Supplementary Table S2. Computed total energy difference between the spin-triplet (S=1) and the spin-singlet (S=0) state of the $X_NY_i$ defects at the PBE and HSE levels of theory. Supplementary Figure S3. Defect Formation energy of 6 $X_NY_i$ dimer defects in h-BN in the nitrogen-poor (a) and the nitrogen-rich (b) condition, respectively, calculated at the PBE level of theory. Supplementary Table S3. Computed binding energy gain of the spin-triplet $X_NY_i$ dimers at the PBE level of theory. Supplementary Figure S4. Ground-state and excited-state structures of the spin-triplet $X_NY_i$ dimer defects computed by using the QE code at the PBE level of theory. Supplementary Figure S5. Many-electron multiplet structures of the $X_NY_i$ dimers at the ground-state (left diagram) and the excited-state geometries computed by using the QET+ED theory. Supplementary Table S4. Computed zero-phonon line (ZPL) and Huang-Rhys (HR) factor of the $X_NY_i$ dimers in h-BN for the $^3A_2 - ^3E$ transition at the PBE level of theory. Supplementary Note S1. Comparison of the VASP and QE results for the ZPL energies and the HR factors. Supplementary Figure S6. Ground-state and excited-state structures of the spin-triplet $X_NY_i$ dimer defects computed by using the VASP code at the HSE level of theory. Supplementary Table S5. Computed anisotropic hyperfine tensors and principal axes of the 6 $X_NY_i$ dimers at the PBE level of theory.

## AUTHOR INFORMATION

**Corresponding Authors**

Hosung Seo − Department of Physics and Department of Energy Systems Research, Ajou University, Suwon, Gyeonggi 16499, Korea; orcid.org/0000-0001-5677-111X;

Email: hseo2017@ajou.ac.kr

**Author Contributions**

J.B. and D.Y. performed the DFT calculations. H.S. planned the study. H.M. and G.G. developed the PyZFS code and the QET code and take charge of QET+ED calculations. H.S. and G.G. supervised the project. All authors contributed to the data analysis and production of the manuscript.

**Notes**

The authors declare no competing financial interest.

**ACKNOWLEDGEMENT**

We thank Yuan Ping for fruitful discussions. This work was supported by the National Research Foundation of Korea (NRF) grant funded by the Korean government (MSIT) (No. 2018R1C1B6008980, No. 2018R1A4A1024157, and No. 2019M3E4A1078666) and by the National Supercomputing Center with supercomputing resources including technical support (KSC-2019-CRE-0154). H.M. and G.G. were supported by AFOSR FA9550-19-1-0358.

**ABBREVIATIONS**

h-BN, hexagonal boron nitride; DFT, density functional theory; QET, quantum embedding theory; ED, exact diagonalization; CBM, conduction band minimum; VBM, valence band maximum; CTL, charge transition level; ZPL, zero-phonon line; DFE, defect formation energy; QE, quantum espresso; ODMR, optically detected magnetic resonance; 2D, 2-dimensional;

**TABLE OF CONTENTS GRAPHIC**

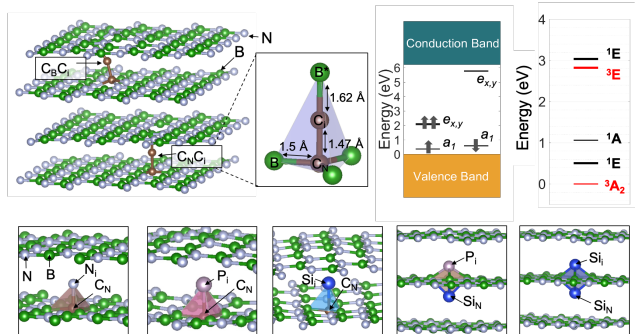